\documentclass[superscriptaddress,twocolumn,showpacs,prl]{revtex4}
\usepackage{graphicx}
\usepackage{tabularx}

\newcommand{\be}{\begin{equation}}
\newcommand{\ee}{\end{equation}}
\newcommand{\beqn}{\begin{eqnarray}}
\newcommand{\eeqn}{\end{eqnarray}}
\newcommand{\nsw}{N_{\mathrm{sweep}}}
\newcommand{\nsa}{N_{\mathrm{samp}}}

\newcommand{\av}{_{\mathrm{av}}}
\newcommand{\chisg}{\chi_{_{\mathrm{SG}}}}

\begin{document}

\title{Absence of an Almeida-Thouless line in Three-Dimensional Spin Glasses}

\author{A.~P.~Young}
\email{peter@bartok.ucsc.edu}
\affiliation{Department of Physics, University of California,
Santa Cruz, California 95064}
                                                                                
\author{Helmut G.~Katzgraber}
\affiliation{Theoretische Physik, ETH H\"onggerberg, 
CH-8093 Z\"urich, Switzerland}

\date{\today}

\begin{abstract}
We present results of Monte Carlo simulations of the three-dimensional
Edwards-Anderson Ising spin glass in the presence of a (random) field.  A
finite-size scaling analysis of the correlation length shows no indication of
a transition, in contrast to the zero-field case.  This suggests that there is
no Almeida-Thouless line for short-range Ising spin glasses.
\end{abstract}

\maketitle

Since the work of Ballesteros {\em et al}.~\cite{ballesteros:00}, there has
been
little doubt that a finite-temperature transition occurs in
three-dimensional spin glasses~\cite{isotropic}. However, the behavior of
a spin glass in a magnetic field is less well understood. In mean field
theory~\cite{edwards:75}, which is taken to be the solution of the
infinite-range Sherrington-Kirkpatrick (SK) model~\cite{sherrington:75},
an Ising system~\cite{vector} has a line of transitions in a magnetic
field~\cite{almeida:78}, known as the Almeida-Thouless (AT) line. This
line separates the paramagnetic phase at high temperatures and fields from
the spin-glass phase at lower temperatures and fields. Although there is
\textit{no} change of symmetry at this transition, the relaxation time
diverges (and for short-range systems so does the correlation length
as we shall see). In the spin-glass phase below the AT line, there is
``replica symmetry breaking'' in which the free energy landscape breaks up
into different regions separated by infinite barriers, and the
distribution of relaxation times extends to infinity.

\begin{figure}
\begin{center}
\includegraphics[width=7.5cm]{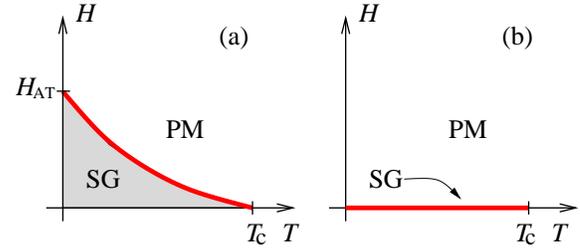}
\end{center}
\vspace*{-0.3cm}
\caption{
(a) $H$--$T$ phase diagram expected according to RSB for the short-range case.
For
$T < T_{\rm c}(H)$
there 
is a spin-glass phase (SG), whereas for $T > T_{\rm c}(H)$ the system is 
in the paramagnetic (PM) state. The value of the critical field at $T=0$ is
called $H_{\rm AT}$. (b) $H$--$T$ phase diagram following the 
predictions from the droplet picture. A spin-glass phase exists only for 
$H = 0$.}
\label{fig:at}
\end{figure}

It is important to know whether the AT line also occurs in more realistic
short-range models, since the two main scenarios that have been proposed
for the spin-glass state differ over this issue.  In the ``droplet
picture''~\cite{fisher:86,fisher:87,fisher:88,bray:86} 
there is
\textit{no} AT line in \textit{any} finite-dimensional spin glass.  By
contrast, the ``replica symmetry breaking'' (RSB)
picture~\cite{parisi:79,parisi:80,parisi:83,mezard:87} postulates that the
behavior of short-range systems is quite similar to that of the
infinite-range SK model which \textit{does} have an AT line as just
mentioned. Both scenarios are illustrated in Fig.~\ref{fig:at}.

Experimentally, it has proved much more difficult to verify
the transition in a field than for
the zero-field transition. For the latter, the
divergence of the nonlinear susceptibility gives clear experimental evidence
of a transition, but unfortunately this divergence no longer occurs in a
magnetic field. Experiments have therefore looked for a divergent relaxation
time, and a careful analysis by Mattsson {\em et al}.~\cite{mattsson:95} finds
that this does \textit{not occur} in a field. However, not all experimental
work has come to the same conclusion~\cite{katori:94}.

In simulations, it is most desirable to perform finite-size scaling (FSS)
on \textit{dimensionless} quantities for reasons that we will discuss
below. One such quantity, the Binder ratio, gave some evidence for the
zero-field transition~\cite{bhatt:85,kawashima:96}. However, the Binder
ratio turns out to be very poorly behaved in a field~\cite{ciria:93b} in
short-range systems, while for the SK model it does indicate a possible
transition~\cite{billoire:03b}, although not with any great precision.
Results of out of
equilibrium simulations on large lattices in four
dimensions~\cite{marinari:98d} were interpreted as evidence for RSB,
although it is not completely clear that the true equilibrium behavior is
probed by this procedure~\cite{barrat:01}.
By contrast, simulations~\cite{takayama:04}
corresponding to experimental protocols in (non-equilibrium)
aging experiments have been
analyzed in terms of a ``dynamical crossover'' consistent with the droplet
picture.

Houdayer and Martin~\cite{houdayer:99} carried out interesting
calculations at $T=0$ to determine $H_{\rm AT}$, the critical field at $T=0$,
see Fig.~\ref{fig:at},
for a simple cubic lattice in
three dimensions. Their results indicated that $H_{\rm AT}=0$, i.e., there is
no AT line, although a subsequent zero-temperature study by Krzakala {\em
et al}.~\cite{krzakala:01} found some evidence of a critical field
for $H \simeq
0.65$, which is much less than the ``mean field'' value for this lattice
\cite{pagnani:03} of around 1.86. However, Krzakala {\em et
al}.~\cite{krzakala:01} could not exclude the possibility that the
critical field is zero.

As noted above, the (dimensionless) Binder ratio has provided some evidence
for zero-field transition at finite-$T$. However, a \textit{much sharper}
signature of the zero-field transition in three dimensions is provided by 
the correlation length~\cite{ballesteros:00} from which a
dimensionless quantity is formed by dividing by the system size $L$. This
approach should also provide evidence \textit{from equilibrium calculations}
for a transition in a field, if one
occurs, and in this paper we use it to determine whether there is an AT line
in a three-dimensional Ising spin glass. Our conclusion will be that there is
not, at least down to fields significantly smaller than the value of 0.65
suggested by Krzakala {\em et al}.~\cite{krzakala:01}.

The Hamiltonian we study is given by
\begin{equation}
{\cal H} = - \sum_{\langle i, j\rangle} J_{ij} S_i S_j - \sum_i h_i S_i
\label{ham}
\end{equation}
in which the Ising spins $S_i = \pm 1$ lie on the sites of a simple cubic
lattice of size $N = L^3$ ($4 \le L \le 12$) with periodic boundary
conditions, and the nearest neighbor interactions $J_{ij}$ are independent
random variables with a Gaussian distribution with mean zero and standard
deviation unity. At each site there is a field $h_i$ which, like the
bonds, \textit{is randomly drawn from a Gaussian distribution}, and whose mean
and standard deviation are given by
\begin{equation}
[h_i]\av = 0, \qquad [h_i^2]\av^{1/2} = H_{\rm r},
\label{field_dist}
\end{equation}
where $[\cdots]\av$ denotes an average over the disorder.
For a symmetric
distribution of bonds, the \textit{sign}
of $h_i$ can be ``gauged away'' so a
uniform field is completely equivalent to a bimodal distribution of fields
with $h_i = \pm H$. Our choice of a Gaussian distribution, which still has
an AT line in mean-field theory, also puts disorder into the
\textit{magnitude} of the $h_i$. We use a Gaussian distribution,
rather than a uniform field,
in order to apply a very helpful test for equilibration, discussed
below.

To determine the correlation length we calculate the wavevector-dependent
spin-glass susceptibility which, for nonzero fields, is defined by
\begin{equation}
\chisg(\mathbf{k}) = \frac{1}{N} \sum_{i, j} \left[\Big(
\langle S_i S_j\rangle_T - \langle S_i \rangle_T \langle S_j\rangle_T
\Big)^2 \right]\av\!\!\!\!\! e^{i\mathbf{k}\cdot(\mathbf{R}_i - \mathbf{R}_j)},
\label{chisg}
\end{equation}
where $\langle \cdots \rangle_T$ denotes a thermal average. As in earlier
work~\cite{ballesteros:00,lee:03} the correlation length of the finite
system is defined to be
\begin{equation}
\xi_L = \frac{1}{2 \sin (k_\mathrm{min}/2)}
\left[\frac{\chi_{SG}(0)}{\chi_{SG}({\bf k}_\mathrm{min})} - 1\right]^{1/2},
\label{xiL}
\end{equation}
where ${\bf k}_\mathrm{min} = (2\pi/L, 0, 0)$ is the smallest nonzero
wavevector.

Now $\xi_L$ satisfies the finite-size scaling form
\begin{equation}
\frac{\xi_L}{L} = \widetilde{X}\left(L^{1/\nu}[T - T_{\rm c}(H_{\rm r})]\right) ,
\label{fss}
\end{equation}
where $\nu$ is the correlation length exponent and $T_{\rm c}(H_{\rm r})$ is the
transition temperature for a field strength
$H_{\rm r}$. Note that there is no power of
$L$ multiplying the scaling function $\widetilde{X}$, as there would be
for a quantity with dimensions. This greatly simplifies the analysis since
the critical point can be seen by inspection as the temperature where data
for different sizes intersect.

On the AT line, $T = T_{\rm c}(H_{\rm r})$, the ``connected correlation 
function''
$\langle S_i S_j\rangle_T - \langle S_i \rangle_T \langle S_j\rangle_T$
becomes long range and so, for an infinite system, the correlation length
and $\chisg(0)$ diverge while for a finite system, $\xi_L/L$ is
independent of $L$. Below the AT line, according to RSB the correlation
functions no longer have a ``clustering property'', i.e.,
$\lim_{|\mathbf{R}_i - \mathbf{R}_j| \to \infty} (\langle S_i S_j\rangle_T
- \langle S_i \rangle_T \langle S_j\rangle_T) \ne 0$, so $\chisg(0)
\propto L^d$ and $\xi_L/L$ increases with $L$. Hence, according to RSB,
the behavior of $\xi_L/L$ should be \textit{qualitatively the same} as at
the zero-field transition, namely it decreases with increasing $L$ above the
transition, is independent of $L$ at the transition, and increases with
increasing $L$ below the transition.

We use parallel tempering to speed up the simulations but
unfortunately it is less efficient in a
field than in zero field~\cite{moreno:03,billoire:03}, because ``chaos''
with respect to a field is stronger than chaos with respect to
temperature.  As a result, the computer time increases very rapidly with
increasing $L$, so it is unlikely that we will be able to study larger
sizes in the near future without a better algorithm. In order to compute the
products of up to four thermal averages in Eq.~(\ref{chisg}) without bias
we simulate four
copies (replicas) of the system with the same bonds and fields at each
temperature.

\begin{figure}
\includegraphics[width=\columnwidth]{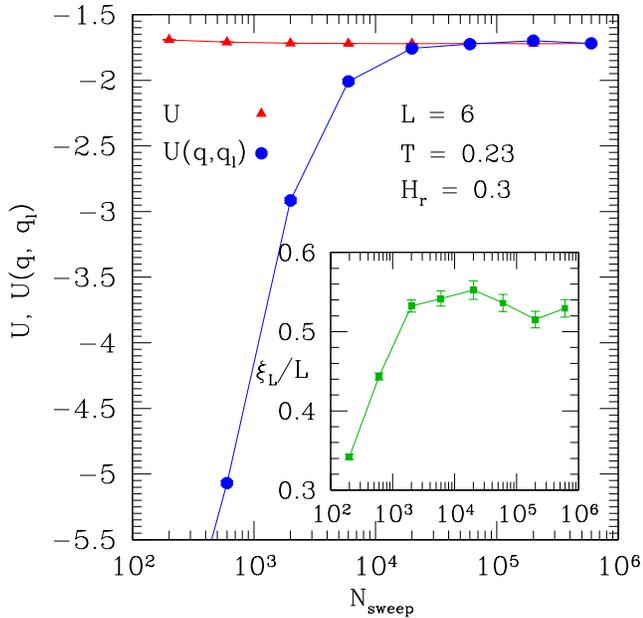}
\vspace*{-0.5cm}
\caption{
An equilibration plot for $L=6, H_{\rm r} = 0.3, T = 0.3$ showing that the
data for the average energy $U$ and the quantity $U(q, q_l)$, defined in
Eq.~(\ref{Uqql}), approach their common equilibrium value from opposite
directions as the number of Monte Carlo sweeps $\nsw$ increases.
The inset shows data for $\xi_L/L$ indicating that it has equilibrated when
$U$ and $U(q, q_l)$ have become equal.
}
\label{equil_all_L6_0.23_0.3}
\end{figure}

Parameters of the 
simulation are shown in Table~\ref{simparams}.
Most of our work is for $H_{\rm r}=0.3$ since this is smaller than the
predicted~\cite{krzakala:01,rand_unif} zero temperature value of
$H_{\rm AT} = 0.65$,
but is not so small that the
results would be seriously influenced by the zero-field transition.
The
lowest temperature is $0.23$ well below the zero-field transition
temperature which is about~\cite{marinari:98} $0.95$.

\begin{table}
\caption{
Parameters of the simulations for $H_{\rm r} = 0.3$. $\nsa$ is the number
of samples, $\nsw$ is the total number of Monte Carlo sweeps for each of
the $4 N_T$ replicas for a single sample, $T_{\rm min}$ is the lowest
temperature simulated, and $N_T$ is the number of temperatures used in the
parallel tempering method. For other values of $H_{\rm r}$, we used the
same parameters but only simulated
$L = 4$, $6$, and $8$.
\label{simparams}
}
\begin{tabular*}{\columnwidth}{@{\extracolsep{\fill}} c r r r l }
\hline
\hline
$L$  &  $\nsa$  & $\nsw$ & $T_{\rm min}$ & $N_{T}$  \\
\hline
  4 & $5000$ & $6.0 \times 10^4$ & 0.23 & 18 \\
  6 & $5319$ & $6.0 \times 10^5$ & 0.23 & 18 \\
  8 & $5000$ & $6.0 \times 10^5$ & 0.23 & 18 \\
 12 & $ 304$ & $6.0 \times 10^7$ & 0.23 & 18 \\
\hline
\hline
\end{tabular*}
\end{table}

For a
Gaussian distributions of bonds \textit{and} fields, the expression for the
average energy, $U = [\langle {\cal H} \rangle_T]\av$, can be integrated
by parts with respect to the disorder distribution, with the result
\begin{equation}
U = U(q, q_l) \, ,
\label{U}
\end{equation}
where
\begin{equation}
U(q, q_l) = \frac{z}{2}\frac{q_l - 1}{T} + \frac{q - 1}{T} H_{\rm r}^2 \, ,
\label{Uqql}
\end{equation}
where $z$ ($=6$ here) is the number of neighbors, $q$ is the spin overlap
given by 
\begin{equation}
q = \frac{1}{N} \sum_{i=1}^N [\langle S_i^{(1)} S_i^{(2)} \rangle_T]\av ,
\label{q}
\end{equation}
and $q_l$ is the ``link overlap'' given by 
\begin{equation}
q_l = \frac{2}{z}\frac{1}{N} \sum_{\langle i, j\rangle}
[\langle S_i^{(1)} S_j^{(1)} S_i^{(2)} S_j^{(2)} \rangle_T]\av .
\label{ql}
\end{equation}
In Eqs.~(\ref{q}) and (\ref{ql}), ``$(1)$'' and ``$(2)$'' refer to two
copies of the system with the same bonds and fields.
Because $U$ will \textit{decrease} as the
system approaches equilibrium and $q$ and $q_l$ will \textit{increase}
(since we initialize the spins in the two copies in random configurations), 
$U$ and $U(q, q_l)$ approach their common equilibrium value
\textit{from opposite directions} and so Eqs.~(\ref{U}) and (\ref{Uqql})
can be used as an
equilibration test. This is 
a generalization to finite fields of a 
test used previously~\cite{katzgraber:01}.
Figure~\ref{equil_all_L6_0.23_0.3} shows
that these expectation are born out. We accept a set of runs as being
equilibrated if $U=U(q, q_l)$ within the error bars. The inset to the
figure shows that $\xi_L/L$ has equilibrated when
$U$ and $U(q, q_l)$ have become equal.

\begin{figure}
\includegraphics[width=\columnwidth]{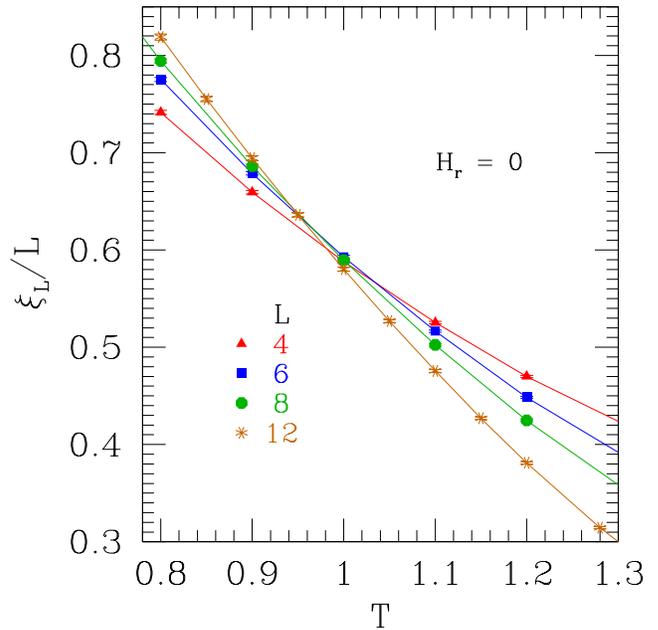}
\vspace*{-0.5cm}
\caption{
Data for $\xi_L/L$ for $H_{\rm r}=0$ for different sizes. Note that there
are clear intersections at roughly a common temperature, with the data
splaying
out at lower temperatures. The temperature of the intersections
is the zero-field
transition temperature, marked $T_{\rm c}$ in Fig.~\ref{fig:at}.
In this data, 5000 samples are used for the disorder average in each system
size.
}
\label{xi_Hr0}
\end{figure}

\begin{figure}
\includegraphics[width=\columnwidth]{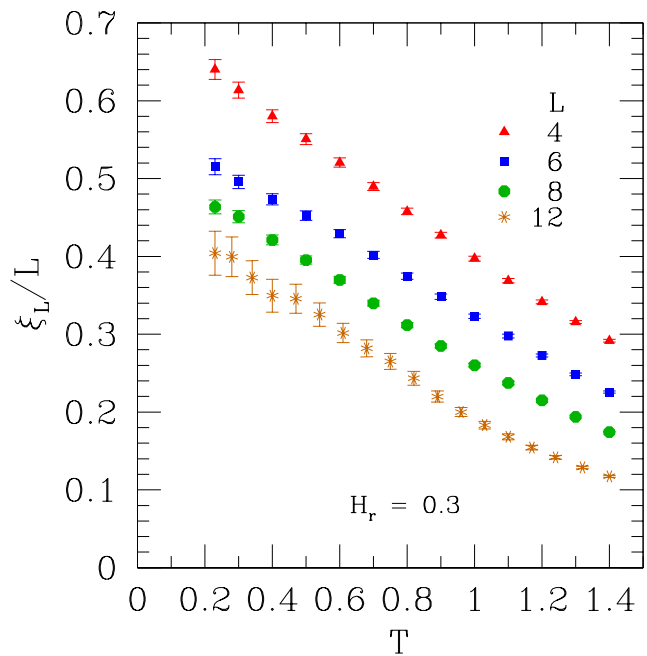}
\vspace*{-0.5cm}
\caption{
Data for $\xi_L/L$ for $H_{\rm r}=0.3$ for different sizes. Note that in
contrast to the zero-field data in Fig.~\ref{xi_Hr0} there is no sign of
intersections down to the lowest temperature $T=0.23$.
}
\label{xi_Hr0.3}
\end{figure}

It is useful to compare results in a field with those at
the zero-field transition. Hence in Fig.~\ref{xi_Hr0} we show data for
$\xi_L/L$ for $H_{\rm r}=0$ for sizes up to $L= 12$.  For these results we set
$\langle S_i \rangle_T = 0$ in Eq.~(\ref{chisg}).  There are clear
intersections, with data splaying out at lower temperatures,
indicating a transition at $T = T_{\rm c} \equiv T_{\rm c}(H_{\rm r}=0)$,
in the region 0.95--1.00,
in agreement with Marinari {\em et al}.~\cite{marinari:98}.

However, the analogous results for $H_{\rm r}=0.3$ shown in
Fig.~\ref{xi_Hr0.3} have no sign of an intersection for sizes up to $L=12$
at temperatures down to $T = 0.23$, which is considerably below the
zero-field transition temperature of about 0.95. This provides quite
strong evidence that there is no AT line, except possibly for fields less
than $0.3$. In order to test this possibility
we have also performed simulations down to
$H_{\rm r} = 0.05$ (for $4 \le L \le 8$),
and again found no intersections.  We have also performed simulations in a
uniform field, finding that data are very similar to those for the random
fields, and have no intersection down to the lowest field studied, $H =
0.1$. 

To go to low fields without passing too close to the zero-field
transition, we also tried a diagonal ``cut'' in the $H_{\rm r}$--$T$ plane
with $H_{\rm r}/T$ kept fixed at the constant value of $0.7$. However, the
equilibration problems were even more severe than for $H_{\rm r}$ fixed at
0.3, and so we have not been able to get useful data for this case.

To conclude, our 
finite-temperature Monte Carlo simulations provide \textit{simple, direct}
evidence from
\textit{equilibrium} calculations
that there is no AT line in three dimensions.
Of course, the numerical data cannot rule out
a transition at exceptionally small fields, or the possibility of a crossover
at much larger sizes to different behavior, but we see no particular
reason for these scenarios to occur.

\begin{acknowledgments}
The work of APY is supported by NSF Grant No.~DMR 0337049. Part of
the simulations were performed on the Asgard cluster at ETH Z\"urich. We
would like to thank F.~Krzakala for a stimulating discussion and for bringing
Ref.~\cite{krzakala:04} to our attention.

\end{acknowledgments}

\vspace*{-0.3cm}

\bibliography{refs,comments}

\begin{thebibliography}{34}
\bibitem[{iso()}]{isotropic}
\bibinfo{note}{There is currently some dispute as to whether there is finite
  temperature transition in a fully isotropic vector spin glass. However, since
  we only consider Ising systems here~\cite{vector} we will not discuss this
  issue.}
\end{thebibliography}

\end{document}